\documentclass[conference]{IEEEtran}
\IEEEoverridecommandlockouts
\usepackage{cite}
\usepackage{amsmath,amssymb,amsfonts}
\usepackage{algorithmic}
\usepackage{graphicx}
\usepackage{textcomp}
\usepackage{xcolor}
\usepackage{soul}

\usepackage{eurosym}
\usepackage{amstext} 
\DeclareRobustCommand{\officialeuro}{%
  \ifmmode\expandafter\text\fi
  {\fontencoding{U}\fontfamily{eurosym}\selectfont e}}

\def\BibTeX{{\rm B\kern-.05em{\sc i\kern-.025em b}\kern-.08em
    T\kern-.1667em\lower.7ex\hbox{E}\kern-.125emX}}
\begin{document}

\title{Linking Parking and Electricity Values to\\Unlock Potentials of Electric Vehicles\\in Portuguese Buildings\\
\thanks{Support for this research was provided by the Fundação para a Ciência e a Tecnologia (Portuguese Foundation for Science and Technology) through the Carnegie Mellon Portugal Program.}
}

\author{\IEEEauthorblockN{Pedro Moura}
\IEEEauthorblockA{\textit{Electrical and Computer Eng.} \\
\textit{University of Coimbra}\\
Coimbra, Portugal \\
pmoura@isr.uc.pt}

\and
\IEEEauthorblockN{Greta K.W. Yu}
\IEEEauthorblockA{\textit{Carnegie Institute of Tech.} \\
\textit{Carnegie Mellon University}\\
Pittsburgh, USA \\
kaiweiy@andrew.cmu.edu}
\and
\IEEEauthorblockN{Shubham Sarkar}
\IEEEauthorblockA{\textit{Carnegie Institute of Tech.} \\
\textit{Carnegie Mellon University}\\
Pittsburgh, USA \\
shubhams@andrew.cmu.edu}
\and
\IEEEauthorblockN{Javad Mohammadi}
\IEEEauthorblockA{\textit{Electrical and Computer Eng.} \\
\textit{Carnegie Mellon University}\\
Pittsburgh, USA \\
jmohamma@andrew.cmu.edu}

}

\maketitle

\begin{abstract}
Large parking lots in public and commercial buildings are increasingly installing the required infrastructure for serving Electric Vehicles (EVs). Utilizing charging and discharging flexibility of parked EVs has the potential to significantly increase the self-consumption of on-site renewable generation and reduce building energy costs. Optimal charging and discharge management of electric vehicles can fill the gap between self-generation and building electric demand while accounting for electricity tariffs. Optimal interactions between EVs and buildings will play a key role in the operation of power networks with a high penetration of distributed energy resources, such as the Portuguese electric grid. However, in Portugal, the existing regulation does not allow financial transactions between buildings and EVs in exchange for charging and discharging. This renders most proposed charging and discharging management strategies impractical. This paper introduces a novel and practical framework to connect electricity and parking values at commercial and public buildings. This framework will manage interactions between building and vehicle in the context of parking time duration and added value services for the charging and discharging periods. The proposed formulation is ready to adopt since it is compatible with the current regulations and relies on existing technologies. The simulation results showcase the cost reduction and self-consumption benefits of the proposed solution for building owners.
\end{abstract}


\section*{Nomenclature}
\subsection*{Inputs}
\addcontentsline{toc}{section}{Nomenclature}
\begin{IEEEdescription}[\IEEEusemathlabelsep\IEEEsetlabelwidth{$V_1,V_2,V_3$}]
\item[$C_P$]  Baseline hourly parking tariff for EVs ($\euro{}/hour$)
\item[$C_C$] EV charging tariff associated with the charging period ($\euro{}/hour$)
\item[$C_D$] EV discharging tariff associated with the discharging period  ($\euro{}/hour$)
\item[$C_F$] Reward to compensate EV charging flexibility ($\euro{}/hour$)
\item[$C_{EE}(h)$] Electricity tariff for power exported to the grid at time step $h$ ($\euro{}/kWh$)
\item[$C_{EI}(h)$] Electricity tariff for power imported from the grid at time step $h$ ($\euro{}/kWh$)
\item[$\mathfrak{t}_c(n)$] Charging period desired by EV owner $n$ during its parking period ($hour$)
\item[$\mathfrak{t}_d(n)$] Maximum discharging period allowed by EV owner $n$ during its parking period ($hour$)
\item[$\mathfrak{t}_P(n)$] Total parking period of EV $n$ ($hour$)
\item[$L(h)$] Net electricity load in the building, excluding the impact of EVs, at time step $h$ ($kW$)
\item[$P_{EV,n}^{+M}$] Maximum charging power for EV $n$ ($kW$)
\item[$P_{EV,n}^{-M}$] Maximum discharging power for EV $n$ ($kW$)
\item[$\eta(n)$] Efficiency of the EV charging/discharging for EV $n$ ($\%$)

\end{IEEEdescription}

\subsection*{Variables}
\addcontentsline{toc}{section}{Nomenclature}
\begin{IEEEdescription}[\IEEEusemathlabelsep\IEEEsetlabelwidth{$V_1,V_2,V_3$}]
\item[$C_E(h)$] Electricity consumption cost of the building at time step $h$ ($\euro{}$)
\item[$C_{EV}(n)$] Cash flow between EV owner $n$ and building during the EV parking period ($\euro{}$)
\item[$t_C(n)$] Net charging period of EV $n$ ($hour$)
\item[$t_D(n)$] Net discharging period of EV $n$ ($hour$)
\item[$P_{EV,n}^+(h)$] Charging power of EV $n$ at time step $h$ ($kW$)
\item[$P_{EV,n}^-(h)$] Discharging power of EV $n$ at time step $h$ ($kW$)
\end{IEEEdescription}

\section{Introduction}
\subsection{Motivation}
The electric power system is quickly changing due to the growing penetration of renewable energy sources in the grid and into buildings. 
The generation mix is increasingly based on distributed, intermittent and non-dispatchable sources and maintaining supply and demand balance is becoming more challenging. 
Simultaneously, the transport sector with Electric Vehicles (EV) is increasingly an important consumer of electricity. However, EVs can be used as controllable loads, using the Grid-to-Vehicle (G2V) system to charge in periods of high renewable generation or low electricity prices \cite{DELGADO2018372}. 
In addition to absorbing power from the grid, EVs can also use some of their storage capacity to inject energy into the grid, using the Vehicle-to-Grid (V2G) system to help ensure the balance between the generation and demand \cite{mohammadi2016fully}. 

In countries with high penetration of hydro and wind power, such as Portugal, there is already a surplus of renewable generation during winter nights. This not only results in technical challenges for grid operators but also leads to low wholesale electricity market prices. With the ambitious targets defined by Portuguese governments, i.e., achieving 100\% renewable electricity generation and 70\% transportation electrification by 2050 \cite{RNC2050}, G2V and V2G play critical roles in providing flexibility for the integration of renewable generation.

Buildings are the foundation and end point of the electric delivery system. In Portugal, the electricity buy-back tariff does not encourage back feeding to the grid. Thus, buildings prefer to increase the self-consumption of local generation. Addressing the mismatch between the on-site renewable generation, e.g., solar photovoltaics (PV), and the electricity demand in buildings requires flexible energy resources. EVs can significantly contribute to providing the much needed flexibility through charging period management, using the Building-to-Vehicle (B2V) system, adjusting the charging period based on renewable generation availability \cite{6861516}. Additionally, using the Vehicle-to-Building (V2B) system, the energy stored in EVs can be injected into the building to compensate periods of low generation (for instance, due to clouds passing by the PV system) or reduce the demand from the grid in periods of high tariffs. 

Large public and commercial buildings, with parking lots, have a high potential to use EVs as a flexible load to increase local renewable generation and demand matching and minimize the electricity costs. Typically EVs and buildings do not belong to the same entity, which means that leveraging EVs' flexibility depends on establishing an economic relationship between building and the EV user. In Portugal, as in most countries, the regulation does not yet allow selling and buying electricity between buildings and entities (e.g., EV user) except the power grid. Therefore, unlocking technical and economic benefits of B2V and V2B requires innovative methods.

\subsection{Related Works}
There is a vast body of works proposing methodologies to implement V2B strategies. Some works have studied V2B in the context of residential buildings. For instance, \cite{Englberger2019} considered a household with PV, EV and storage with the goal to minimize the operational costs. However, in residential buildings, the EV and the building belong to the same entity. Therefore, there are no economic transactions between building and EV.

There are also several works focused on the integration of EVs in office buildings. Authors in \cite{CAO2019113347} have assessed the impact of EVs on the integration of renewable generation in office buildings. Moreover, \cite{JIN2017480} has considered an office building with flexible demand and EVs with the objective of ensuring load leveling. Also, \cite{THOMAS20181188} considered an office building equipped with PV, storage and EVs that aims to minimize the total energy costs. The specific case of University campuses is also considered in several works. For example, \cite{8550711} considered a University campus with PV, battery and EVs with the objective of minimizing peak load while \cite{IOAKIMIDIS2018148} has studied a University building with PV systems and EVs with the objective of maximizing peak-shaving and valley-filling. These papers do not take into account the need for establishing an economic relationship between building and EVs although they belong to different entities.

The economic relationship between EV users and buildings is explored by some researchers. Authors in \cite{en11082165} considered an office building set up with PV and EVs with the objective of minimizing energy costs. Also, \cite{KUANG2017427} considers a building with renewable generation and storage and EVs charging directly with the generated energy with the objectives of minimizing costs and greenhouse gas emissions. These literature assume that buildings and EVs can exchange electricity in return for money which does not comply with existing regulation in most countries.

\subsection{Contribution}
The main contribution of this work is connecting the value of electricity and parking duration and using this connection to design a novel V2B methodology that can be implemented with the existing regulation in Portugal (where the electricity trade between buildings and EVs is not allowed). The cornerstone of the proposed methodology is the parking time and cost. This solution enables the buildings to sell parking spaces instead of electricity. In this context, EV charging would be considered as an added-value service. In other words, EV users will pay based on the duration of the charging period in the same way as he is paying for parking periods. Similarly, the parking cost can be reduced if the EV user allows the use of the energy stored in the EV battery to be injected into the building. Such energy is only going to be used in the building and not for V2G. Therefore, there is no need to change the business relationship with the grid and the building is not trading electricity, but only providing parking services.

The tariff of the charging and discharging services should accommodate higher discharging tariffs to provide incentives and compensate for the likely battery degradation. Additionally, in order to incentivize higher charging flexibility, a reward is used for the idle period (i.e., parking period without charging or discharging), since longer idle periods allow more scheduling flexibility of charging and discharging. With this strategy, the EV user only has to provide information about the desired parking period, the desired charging period and the maximum allowed discharging period. This eliminates the need for data that is not easily obtained (such as the State of Charge) and simplifies the system implementation.

\subsection{Paper Organization}
The remainder of the paper is structured as follows. Section~II presents the problem formulation. Section~III presents the test case and simulation results. Finally, Section~IV summarizes the paper, emphasizing its main conclusions.

\section{Problem Formulation}
\subsection{Objective Function}
The proposed problem aims at minimizing the costs in the building over all time steps (i.e., $H$). The objective function \eqref{eq:obj} accounts for building electricity costs, as well as cash flow associated with the parking, charging and discharging of $N$ EVs parked at that building.

\small
\begin{equation}
\textrm{min} \sum_{h=1}^{H}C_E(h)-\sum_{n=1}^{N}C_{EV}(n) \label{eq:obj}
\end{equation}
\normalsize

\noindent The first term in \eqref{eq:obj} represents the net cost of building electricity consumption and self-generation, which is detailed in \eqref{eq:ce}. This net cost consists of two parts: (i) the cost of energy drawn from the grid; and (ii) the financial compensation for the energy injected into the grid. In \eqref{eq:ce}, the first term represents the case that the building has electricity deficit and it has to import power from the grid. The second term captures the case that the building has electricity surplus and exports power back to the grid. Note, $\mathbb{P}[.]_{+}$ is the operator that preserves positive values and equates non-positive values to zero.
\small
\begin{multline}
C_E(h)=\mathbb{P}\left[ L(h)+\sum_{n=1}^{N}P_{EV,n}^+(h)-\sum_{n=1}^{N}P_{EV,n}^-(h) \right]_{-} \cdot C_{EI}  \cdot \Delta h 
\\
+\mathbb{P}\left[ L(h)+\sum_{n=1}^{N}P_{EV,n}^+(h)-\sum_{n=1}^{N}P_{EV,n}^-(h) \right]_{+} \cdot C_{EE} \cdot \Delta h \label{eq:ce}
\end{multline}
\normalsize

\noindent The second term of \eqref{eq:obj} shows the financial transactions between EVs and the building. $C_{EV}(n)$ is comprised of multiple time based components. The time references that are submitted by the EV owner $n$ before entering the parking lot include: (i) $\mathfrak{t}_P(n)$, requested parking period (ii), $\mathfrak{t}_c(n)$, requested charging period (iii), and $\mathfrak{t}_d(n)$, allowed discharging period. 
\small
\begin{multline}
C_{EV}(n)\hspace{-.6mm}=\hspace{-2mm}\overbrace{\mathfrak{t}_P(n)}^{\textrm{Parking period}}\hspace{-1mm}\cdot C_P  + \overbrace{t_{C}(n)}^{\textrm{Charging Period}}\hspace{-2mm}\cdot C_C\hspace{-1mm} + \overbrace{t_D(n)}^{\textrm{Discharging period}}\hspace{-2.5mm}\cdot C_D\hspace{-2mm}\\
+\underbrace{\left(\mathfrak{t}_P(n)-t_{C}(n)-t_D(n)\right)}_{\textrm{Idle period}}\cdot C_F\label{eq:ct}
\end{multline}
\normalsize

\noindent The total parking costs, for each EV $n$, depend on the parking period, charging period and discharging period, as well as associated tariffs. Equations \eqref{eq:tc} and \eqref{eq:td} derive these periods shown in \eqref{eq:ct}. Note, function $\mathbb{C}$ will count the number of non-zero elements, hence, \eqref{eq:tc} and \eqref{eq:td} calculate total charging and discharging periods of individual vehicles.

\small
\begin{equation}
t_{C}(n)=\Delta h \sum_{h=1}^{H} \mathbb{C}\left[P_{EV,n}^+(h)\right]
\label{eq:tc}
\end{equation}

\begin{equation}
t_{D}(n)=\Delta h \sum_{h=1}^{H} \mathbb{C}\left[P_{EV,n}^-(h)\right]
\label{eq:td}
\end{equation}
\normalsize

\subsection{Constraints}
The discussed objective is subject to constraints such as parking, charging and discharging time limitations, as well charging/discharging power constraints. The charging duration achieved until the end of the parking period 
should be enough to ensure the satisfaction of the required total charging period requested by the user and to compensate the discharging period \eqref{eq:tct}. Note, the compensation accounts for the discharge and losses caused by battery discharging and since the requested charging period was defined base on the maximum power, the charging and discharging periods must be corrected by the ratio between the average  and the maximum power.

\small
\begin{equation}
t_C(n)=\overbrace{\mathfrak{t}_c(n)}^{\textrm{Requested charging period}}\frac{P_{EV,n}^{+M}}{\overline{P}_{EV,n}^+}+\frac{t_D(n)}{\eta(n)}\frac{ \overline{P}_{EV,n}^-}{P_{EV,n}^{-M}} \label{eq:tct}
\end{equation}
\normalsize

\noindent The discharging period must be lower than the maximum discharging period allowed by the user \eqref{eq:tdm1} and than the charging period \eqref{eq:tdm2}, in order to ensure that a state of charge lower than the initial value is never achieved (since the SoC is not monitored, the risk of reaching the minimum SoC is eliminated). 

\small
\begin{equation}
t_D(n) \leq \overbrace{\mathfrak{t}_d(n)}^{\textrm{Allowed discharging period}} \label{eq:tdm1}
\end{equation}
\normalsize

\small
\begin{equation}
t_{D}(n) \leq t_{C}(n) \label{eq:tdm2}
\end{equation}
\normalsize

\noindent In order to allow the charging and discharging, the parking, charging and discharging periods must be positive \eqref{eq:tp}. 

\small
\begin{equation}
\mathfrak{t}_P(n)>0, \;\ \mathfrak{t}_C(n)>0, \;\ \mathfrak{t}_D(n)>0 \label{eq:tp}
\end{equation}
\normalsize

\noindent Additionally, the charging \eqref{eq:ch-limit} and discharging \eqref{eq:dc-limit} power is limited by the charging infrastructure. 

\small
\begin{equation}
0\leq P_{EV,n}^+(h) \leq P_{EV,n}^{+M} \label{eq:ch-limit}\\
\end{equation}

\begin{equation}
0\leq P_{EV,n}^-(h) \leq P_{EV,n}^{-M} \label{eq:dc-limit}\\
\end{equation}
\normalsize
\
\section{Simulation Results}
\subsection{Data and Test Cases}
The test cases use data from the Department of Electrical and Computer Engineering at the University of Coimbra (Portugal). The building has a total area of about 10,000~$m^2$, an electricity consumption of about 500~$MWh/year$ and a PV system sized at 80~$kWp$, which covers about 16\% of the existing electricity demand \cite{Paula2018}. The PV generation was adjusted for a future scenario where it addresses 50\% of the electric demand, in order to ensure periods with renewable generation surplus to be used in the test cases.

\begin{figure}[htbp]
\centerline{\includegraphics[trim={0 0.5cm 0.5cm 0.5cm},width=0.5\textwidth] {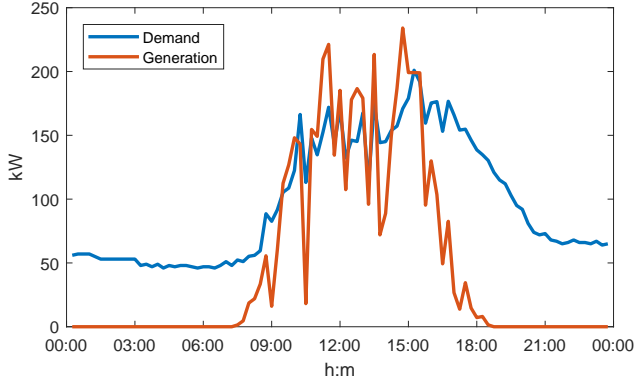}}
\caption{Generation and demand profile for a weekday in March}
\label{fig1}
\end{figure}

One month of intermediate consumption and generation was considered, being elected a weekday in March. Fig.~\ref{fig1} presents the considered generation and demand profiles. The tariff for the electricity imported from the grid is 78.3~$\euro/MWh$ in the super off-peak (0h - 2h), 84.3~$\euro{}/MWh$ in the off-peak (2h - 6h, 22h30 - 24h), 121~$\euro{}/MWh$ in the half-peak (10h30 - 17h, 19h30 - 22h30) and 188.8~$\euro{}/MWh$ in the peak periods (8h - 10h30, 17h - 19h30), and the tariff for the energy injected into the grid is 35.8~$\euro{}/MWh$ in all periods. For the parking tariffs, it was considered 0.5~$\euro{}/hour$, 2~$\euro{}/hour$, 3~$\euro{}/hour$, respectively, for the parking, charging and discharging periods, as well as 0.5~$\euro{}/hour$ for the flexibility reward. The used chargers have a maximum charging/discharging power of 10~$kW$ and 93\% of efficiency. In the assessment, two test cases were considered. Case~1 with a variable number of EVs and a similar profile in terms of charging, discharging and parking requirements, and Case~2 presenting a constant number of EVs and different requirements. 

\subsection{Results}
In Case~1, the variation of the total costs and impact on the net load were assessed, considering an increasing number of EVs, ranging from 0 to 20. In such a scenario, similar assumptions of parking, charging and discharging periods were used for all EVs. It was considered a parking period between 8~$a.m.$ and 8~$p.m.$, aligned with the typical use of office spaces and parking, charging and discharging periods with an average value of 8~$hours$, 2~$hours$ and 0.75~$hours$, respectively. 

\begin{figure}[htbp]
\centerline{\includegraphics[trim={0 0.5cm 0.5cm 1cm},width=0.5\textwidth] {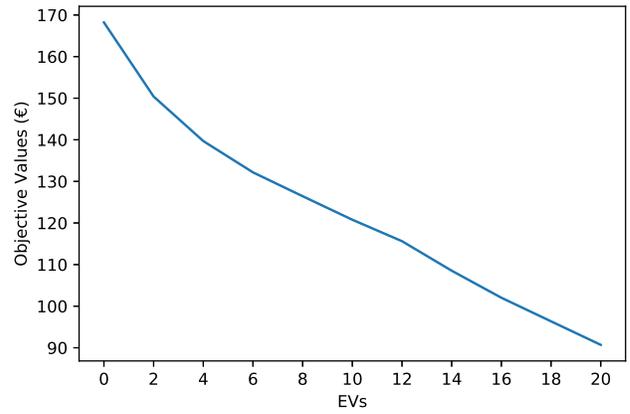}}
\caption{Variation of the objective function with an increasing number of EVs}

\label{fig2}
\end{figure}

Fig.~\ref{fig2} presents the variation of the objective function with the increasing number of EVs. As can be seen, an increasing number of EVs always lead to a decrease in costs. This is the result of the defined price scheme, since the charging price was designed to be always higher than the cost of the energy required to ensure such charging and the discharging price was designed in order to ensure that a charging/discharging cycle presents a lower cost than the difference between the electricity consumed and injected into grid.   

\begin{figure}[htbp]
\centerline{\includegraphics[trim={0 0.5cm 0.5cm 0.8cm},width=0.5\textwidth] {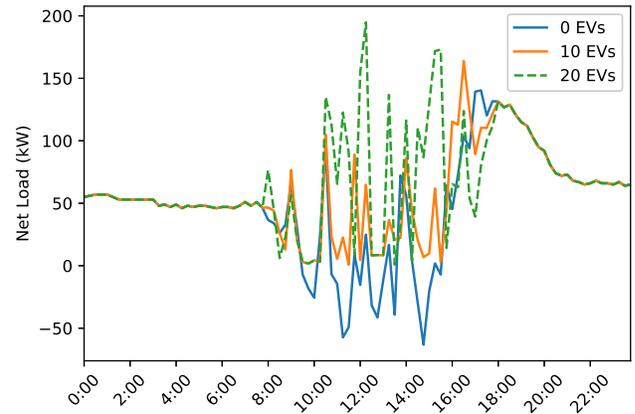}}
\caption{Variation of the net load with an increasing number of EVs}
\label{fig3}

\end{figure}

Fig.~\ref{fig3} presents the variation of the net load in the building with an increasing number of EVs. By increasing the number of EVs until 8 the negative net load is avoided. However, after compensating the generation surplus, an increasing number of EVs (higher than 16) will lead to an increase of the peak power, and to a quick variation on the net load, being fundamental a higher level of charging flexibility to smooth the net load profile.

In Case~2, different charging, discharging and parking periods were considered in order to assess the impact of different charging and discharging profiles. Instead of comparing different quantities of EVs, this scenario considers different requirements in terms of parking and charging. Therefore, a constant number of 10 EVs is considered. The considered parking, charging and discharging periods vary between 0.5 and 10~$hours$, 0.5 and 2.25~$hours$, and 0 and 1.5~$hours$, respectively, presenting an average of 5.2, 1.4 and 0.9~$hours$ for the parking, charging and discharging, respectively. 

\begin{figure}[htbp]
\centerline{\includegraphics[trim={0 0.5cm 0.5cm 1cm},width=0.5\textwidth] {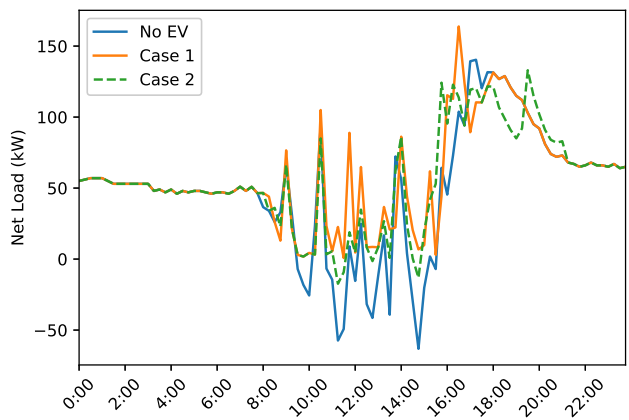}}
\caption{Variation of the net load with Case 1 and Case 2}
\label{fig4}
\end{figure}

Fig.~\ref{fig4} presents a comparison between the net load achieved with Case~2, Case~1 and without EVs. In both scenarios, the number of EVs was 10, but they have similar requirements in Case~1 and differed requirements and a lower average in Case~2. The achieved value of the objective function in Case~2 (127.06~$\euro{}$) was higher than in Case~1 (120.78~$\euro{}$). In Case~1 all the negative net load periods were avoided, therefore ensuring the self-consumption of all generated energy, which was not possible in Case 2. Such results are the consequence of the lower flexibility (short parking and charging periods) in Case~2, therefore limiting the options to use the charging and discharging of EVs to compensate for the mismatch between the local generation and demand and to concentrate the charging in periods with lower tariffs.

\begin{figure}[htbp]
\centerline{\includegraphics[trim={0 0.5cm 0 0.8cm},width=0.5\textwidth] {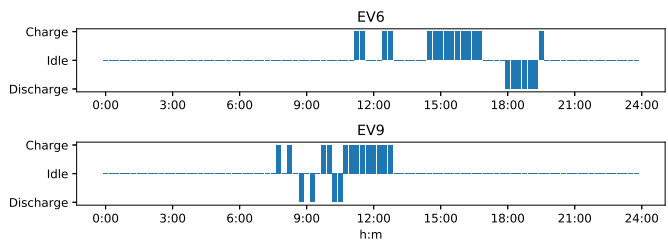}}
\caption{Charging/discharging profiles for EVs with different parking period}
\label{fig5}
\end{figure}

In Case 2 the charging and discharging profiles of different EVs can be compared. Fig.~\ref{fig5} compares the charging and discharging of two EVs requiring the same charging period and allowing the same discharging period, but with very different parking periods (12~$hours$ for EV6 and 5~$hours$ for EV9). Despite the same charging period, a longer parking period allows EV6 to start the charging latter and to concentrate the discharging in a period of lower generation level. EV9 alternates between periods of charging and discharging, being used the discharging to compensate periods in which the net load is momentarily lower. However, the lower parking period does not enable the concentration of discharging in periods with more impact on the net load and costs. The switching between charging and discharging periods demonstrates the capability to ensure that a SoC lower than the initial value is never reached, therefore avoiding the minimum SoC. Such results also demonstrate the need for flexibility in order to ensure the optimum management of the charging and discharging process to ensure the objectives from the building and user point-of-views, therefore proving the relevance of rewarding the flexibility provided by each EV.

\section{Conclusion}
This paper proposes a novel methodology to solve the limitations for the implementation of V2B systems in Portugal. Specifically, this paper proposes a mathematical formulation that enables V2B interactions while following the current Portuguese regulation that prohibits financial transaction to purchase and sell electricity between buildings and EVs. This novel paradigm is based on the value of parking time duration and added value services for the charging and discharging periods. The proposed formulation considers cost minimization from the building point-of-view while accounting for electricity and parking tariffs. 

The formulation was simulated for the University of Coimbra testbed, considering two case studies; similar EV specifications and non-similar EV requirements. The simulation results show that the proposed solution results in an increased contribution of EVs to minimize the building electricity costs, as well as increasing the self-consumption of the on-site renewable generation. This is particularly important in Portugal since the electricity buy-back tariffs are low. The presented results also demonstrated the crucial role of the flexibility (i.e., longer parking time duration than the required charging periods) in facilitating the charging/discharging management process and providing economic and technical benefits to the building.

\vspace{2mm}

  \bibliographystyle{./bibliography/IEEEtran}
  \bibliography{./bibliography/IEEEabrv,./bibliography/IEEEexample}

\end{document}